\documentclass[aps,prl,preprint,superscriptaddress,showpacs]{revtex4}
\usepackage{graphicx}
\usepackage{dcolumn}
\usepackage{amsmath}
\usepackage{epsfig}
\begin{document}
\title{Direct resolution of unoccupied states in solids via two photon photoemission}
\author{W. Schattke}
\affiliation{
Institut f\"ur Theoretische Physik und
Astrophysik der Christian-Albrechts-Universit\"at zu Kiel,
Leibnizstra{\ss}e 15, 24118 Kiel}
\affiliation{
Donostia International Physics Center DIPC, P. Manuel de Lardizabal 4,
20018 San Sebasti\'an, Spain}
\author{E.E. Krasovskii}
\affiliation{
Institut f\"ur Theoretische Physik und
Astrophysik der Christian-Albrechts-Universit\"at zu Kiel,
Leibnizstra{\ss}e 15, 24118 Kiel}
\author{R. D\'{i}ez Mui\~{n}o}
\affiliation{
Donostia International Physics Center DIPC, P. Manuel de Lardizabal 4,
20018 San Sebasti\'an, Spain}
\affiliation{Centro de F\'{\i}sica de Materiales Centro Mixto
CSIC-UPV/EHU,
Apartado 1072, 20080 San Sebasti\'an, Spain}
\author{P. M. Echenique}
\affiliation{
Donostia International Physics Center DIPC, P. Manuel de Lardizabal 4,
20018 San Sebasti\'an, Spain}
\affiliation{Centro de F\'{\i}sica de Materiales Centro Mixto
CSIC-UPV/EHU,
Apartado 1072, 20080 San Sebasti\'an, Spain}
\affiliation{Departamento de F\'{\i}sica de Materiales,
Facultad de Qu\'{\i}micas,
UPV/EHU, Apartado 1072, 20080 San Sebasti\'an, Spain}

\begin{abstract}
Non-linear effects in photoemission are shown to open a new access
to the band structure of unoccupied states in solids, totally
different from hitherto used photoemission spectroscopy. Despite its
second-order nature, strong resonant transitions occur, obeying
exact selection rules of energy, crystal symmetry, and momentum.
Ab-initio calculations are used to demonstrate that such structures
are present in low-energy laser spectroscopy experimental
measurements on Si previously published. Similar resonances are
expected in ultraviolet angle-resolved photoemission spectra, as
shown in a model calculation on Al.
\end{abstract}

%
%


\maketitle

Our understanding of the electronic properties of materials has
spectacularly progressed in the last decade. The spread of
theoretical methods based on first principles, together with the
development of novel and sophisticated experimental techniques, have
led to accurate descriptions of the ground state in many condensed
matter systems. In particular, experimental techniques based on the
photoemission of electrons from the sample currently are the
dominant tool to understand solids and lower dimensional systems in
terms of their electronic structure.

In contrast, a major challenge remaining in condensed matter physics
is the description of excited electronic states at the same level of
accuracy reached in the description of the ground state. From the
experimental point of view, information is limited: A momentum
resolved direct band structure spectroscopy for unoccupied states is
available only through inverse photoemission spectroscopy (IPS)
\cite{FausterIPS,SkibowskiIPS}. Despite a long history, IPS has
never attained an accuracy comparable to that of angle-resolved
photoemission (ARPES) \cite{Buch,Huefner95} for occupied states.
Other methods, such as electron diffraction at very low energies
\cite{Strocov}, are rather indirect. Furthermore, these methods are
affected by the break of three dimensional symmetry brought forth by
the surface: A direct access to the band structure of the solid is
thus prevented because of the cumbersome momentum non-conservation.

However, a new generation of broadly tunable light sources of high
intensity has been developed and can dramatically alter such
situation. Multiphoton processes can be the key to study electronic
excited states. Investigation of localized states, as e.g. surface
and image states, with two-photon excitation has already attracted
much interest ~\cite{Faustermono}. Pump-probe techniques have been
extensively employed for the study of electron and adsorbate
dynamics \cite{Hofer97,Petek00}. The spectroscopy of bulk electronic
structure of the continuum by two-photon processes has received
comparatively little attention
\cite{Shudo,Kentsch,Banfi,Petek,Kirschner}. The role of intermediate
unoccupied states in these transitions has not been fully understood
yet. Theoretical models on this issue mainly rely on density matrix
time evolution \cite{Hertel,Ueba} and non-linear response treatments
\cite{Bennemann}. In a full theory, the dynamical properties of the
system should be described at the same level of accuracy provided by
the one-step model of one-photon photoemission based on {\it
ab-initio} calculations. The work that we present in this Letter is
a necessary first step along this direction.

We here develop and justify a band-mapping procedure for unoccupied
states in bulk, which surpasses the inaccuracies and drawbacks of
the standard traditional mapping procedures. It is based on
non-linear effects appearing in the photoemission process and on the
subsequent appearance of non-linear resonant peaks in the
photoemission spectra. Because of their narrowness, these peaks
should be easier to detect with tunable light sources of low
bandwidth. Starting from the one-step model, we show that the
monochromatic two-photon photoemission yields directly the energy
difference between an occupied state and an unoccupied state, within
the theoretical accuracy of the linewidths of the participating
states. The momentum identification is unique in surface parallel as
well as in perpendicular direction for the initial to intermediate
transition. The bands involved in the transition are assigned
through symmetry selection. The range of application of the scheme
is far reaching, as it opens a new access to the band structure of
unoccupied states in solids. To prove our point, we analyze here two
different sets of experimental photoemission spectra in Si
\cite{Shudo,Kentsch}, and bring them together under one common
interpretation.

We theoretically investigate contributions to the photoemission
spectra of orders higher than linear in the perturbation expansion
with respect to the photon intensity. We restrict the considerations
here to the lowest order correction and obtain the probability per
unit of time of accepting an electron in the detector (i.e., the
photocurrent $J=J_{1phot}+J_{2phot}$) for transitions from a
definite initial state $|a\rangle$ to a final state $|f\rangle$. The
latter is specified by energy $E^f_k$ and momentum $k$. It has to be
summed over occupied initial single particle states (momentum
$k_{a}$ and energy $E_{k_a}^a$) and unoccupied intermediate states
($k_{z}$ and $E_{k_z}^z$).
\begin{eqnarray}\label{prob0}
&J& \propto \sum_{k_a\le k_F}
\left|h_{fa}\right|^2\delta(E_k^f-E_{k_a}^a-\hbar\omega)
\nonumber \\
&+&\frac{1}{4}\sum_{k_a\le
k_F}\delta(E_k^f-E_{k_a}^a-2\hbar\omega)\left|\sum_{k_z}
\frac{h_{fz}h_{za}}{E_{k_z}^z-E^a_{k_a}-\hbar\omega-i\eta}\right|^2
\end{eqnarray}
To obtain the above formula, real one-photon second-order emission
processes are disregarded.  The transition matrix elements are
represented by $h_{ik}$. The sum runs over intermediate states
$|z\rangle$ reached by virtual one-photon transitions from the
initial state observing Pauli exclusion principle. Compared with
bulk band structure investigations by one-photon photoemission such
transitions are very special because full three-dimensional momentum
is conserved\cite{footnote}. It is this property which promises a
large step forward in photoemission spectroscopy. First, because it
allows the access to unoccupied states. Second, because it reaches
the maximum resolution provided by first principles theories.

The meaning of $\eta$ in Eq.~(\ref{prob0}) is two-fold: First, it
represents the decay width of the intermediate state as a lifetime
and the dephasing of the transition amplitude owing to intrinsic
elastic and inelastic processes ($\eta_0$). Second, the strong
external photon field may broaden the energies by its ponderomotive
force in high-intensity light sources ($\eta_1$). Thus,
$\eta=\eta_{0}+\eta_{1}$ makes $J$ finite for all $\eta$ in
Eq.~(\ref{prob0}). The ratio $R$ between the two-photon and
one-photon currents can be approximated by
$R(E_0)=\left|\frac{eE_0a}{\eta}\right|^2$, with $a$ being the
lattice constant, $e$ the electron charge, and $E_0$ the external
field.

From the expression above, we can get the order of magnitude of the
two-photon vs. one-photon current $R$. The decay width $\eta$ is
below 1 eV for intermediate states near the vacuum level. Typical
values of $E_0$ for Ti:sapphire laser systems range from $\approx
10^{6}$ V/m~\cite{Kirschner} to $\approx 10^{10}$
V/m~\cite{Irvine,Miaja}. The corresponding ratio is $R =
0.2\times10^{-6}/\eta^2$ to $R = 0.2\times10^{2}/\eta^2$
respectively (with $\eta$ in eV), and higher for a free electron
laser.  The latter is at the borderline of perturbation theory
validity. Hence, even for the smaller laser intensity, the ratio of
two-photon to one-photon intensity is still significant for small
broadening widths of some meV. In the following, and for the sake of
a working order of magnitude, we take $R=0.2\times10^{2}/\eta^2$ and
use $\eta$ as input parameter.

As a first application of our formalism, we focus into photoemission
processes triggered by strong optical laser fields in the
femtosecond range. We address here two-photon photoemission
processes from the technologically significant Si(001) semiconductor
surface. Resonant photoexcitation in this surface has been already
observed experimentally~\cite{Shudo,Kentsch}. In
Ref.~\onlinecite{Shudo}, a two-photon bulk transition was
identified. The strong enhancement of the photocurrent, with a peak
width of 0.3 eV ~\cite{Shudo}, must be attributed entirely to an
intermediate state. Furthermore, a strong peak from a two-photon
process is observed in Ref.~\onlinecite{Kentsch} as a transition
from the $\Delta_{2'}$ valence band to the $\Delta_{5}$ conduction
band. This structure could be a candidate for the transitions
discussed here as well. In any case, we remark that isolated single
resonances are not required for our analysis. In dispersing bands,
the required energy matching between the photon energy and each one
of the electron transitions is not rare. This will increase the
number of cases with matching conditions, as well as a a broadening
of the peaks.

\begin{figure}[hb]
\psfig{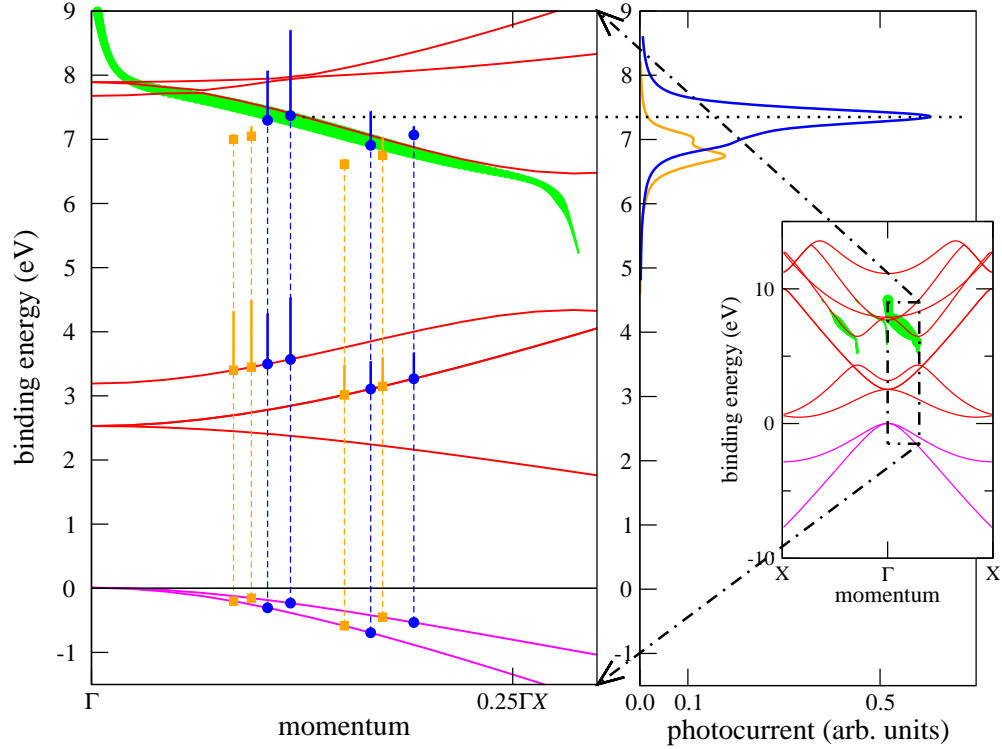} \caption{(Color
online) Two-photon photocurrent and some relevant two-photon
transitions in Si(001). Right panel: Photocurrent vs. binding
energy. Yellow (blue) spectrum corresponds to 3.6 (3.8) eV photon
energy. The full band structure is shown in the inset as well.
Thickened green lines denote LEED states in the complex band
structure and the thickness gives the coupling strength to the
vacuum of that band. Left panel: Magnified view of a reduced portion
of the band structure that contributes to the photocurrent. The
transitions are indicated by squares and circles for 3.6, 3.8 eV
photon energy, respectively. LEED states denoted as in right panel.
Vertical solid lines rising from the squares and circles show the
magnitude of the coupling matrix element for this transition.}
\label{bstr}
\end{figure}

We first calculate the Si band structure, which is plotted in
Fig.~\ref{bstr}, using density functional theory (DFT) and the
augmented plane wave (APW) formalism \cite{Krasovskii07}. The
two-photon photocurrent, also shown in Fig.~\ref{bstr} for two
different values of photon energy, is obtained from this DFT
calculation as well. Initial and intermediate states are scattering
states, which can be represented by Bloch waves (bulk wave
functions) incident from the interior of the crystal and scattered
by the surface. In addition, surface states, which are not related
to any bulk solutions may appear among initial and intermediate
states. Because we are interested only in the asymptotics of the
final state at the detector, the summation over the final states
reduces to the calculation of the time-reversed LEED function,
similar to the first order one-step theory. The LEED state is
calculated ab-initio assuming a step-like surface barrier and an
optical potential $V_i$. In calculating the squared modulus of the
matrix element $h_{za}$, we sum up incoherently over all Bloch
constituents, thus assuming an absolutely rough surface. The matrix
elements $h_{zf}$ are assumed to depend only on the momentum
difference $k_f-k_z$, through a Lorentzian function.

\begin{figure}
\includegraphics[width=0.80\textwidth]{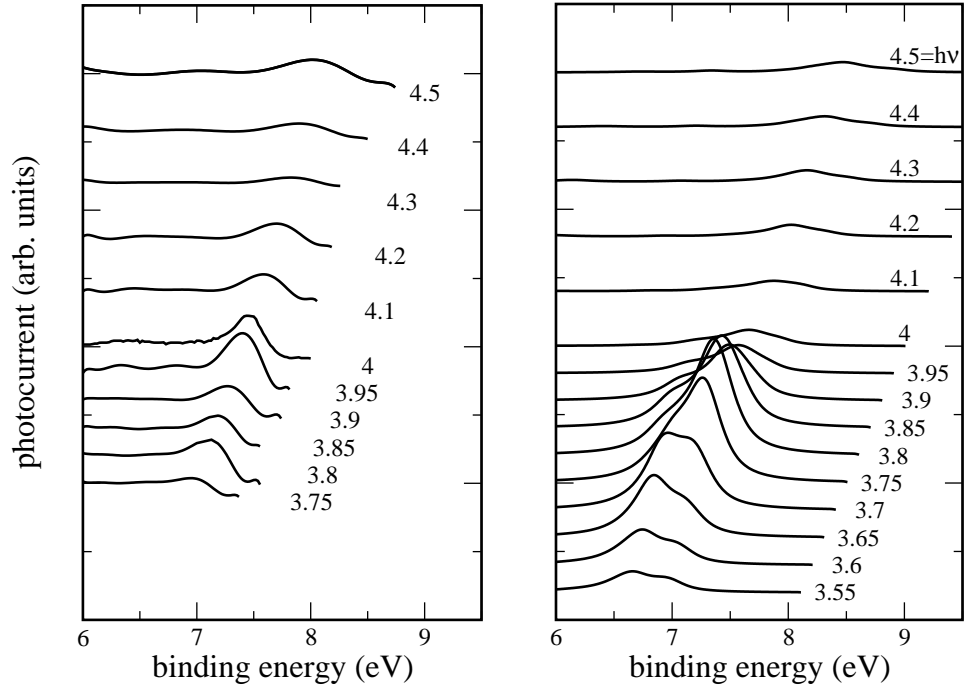}
\caption{Two-photon spectra from Si(001). Left panel: Experimental
values~\cite{Shudo} for $\hbar\omega$ = 3.75 to 4.5 eV. Right panel:
one-step calculation for $\hbar\omega$ = 3.55 to 4.5 eV, with
broadening $\eta=0.15$ eV for initial and intermediate states, and
optical potential $V_i=0.125$ eV.} \label{sispectra}
\end{figure}

Figure \ref{bstr} shows that distinct peaks in the two-photon
photocurrent are associated with initial to intermediate state
resonances, corresponding to momentum conserving transitions. In
addition, the spectra show larger enhancements when the final state
is a strong current carrying wave. This is further illustrated by
the ab-initio calculations of the two-photon photocurrent for the
Si(001) surface, shown in Fig.~\ref{sispectra}. The theoretical
photocurrent of Fig.~\ref{sispectra} very much resembles that
measured in Ref.~\onlinecite{Shudo}, assuming an overall shift of
the theoretical unoccupied states by $0.3$ eV to higher energy,
within the picture of a scissor operator. Applying the scissor
operator, one can see that the maximum of the photocurrent peaks
shifts by approximately 0.15 eV, to photon energy of 3.95 eV, as in
experiment. The binding energies given by the peaks themselves then
align, too. This proves the possibility to precisely adjust the
energies to experiment.

\begin{figure}
\includegraphics[width=0.75\textwidth,clip=true,trim= 0 0 0 7]{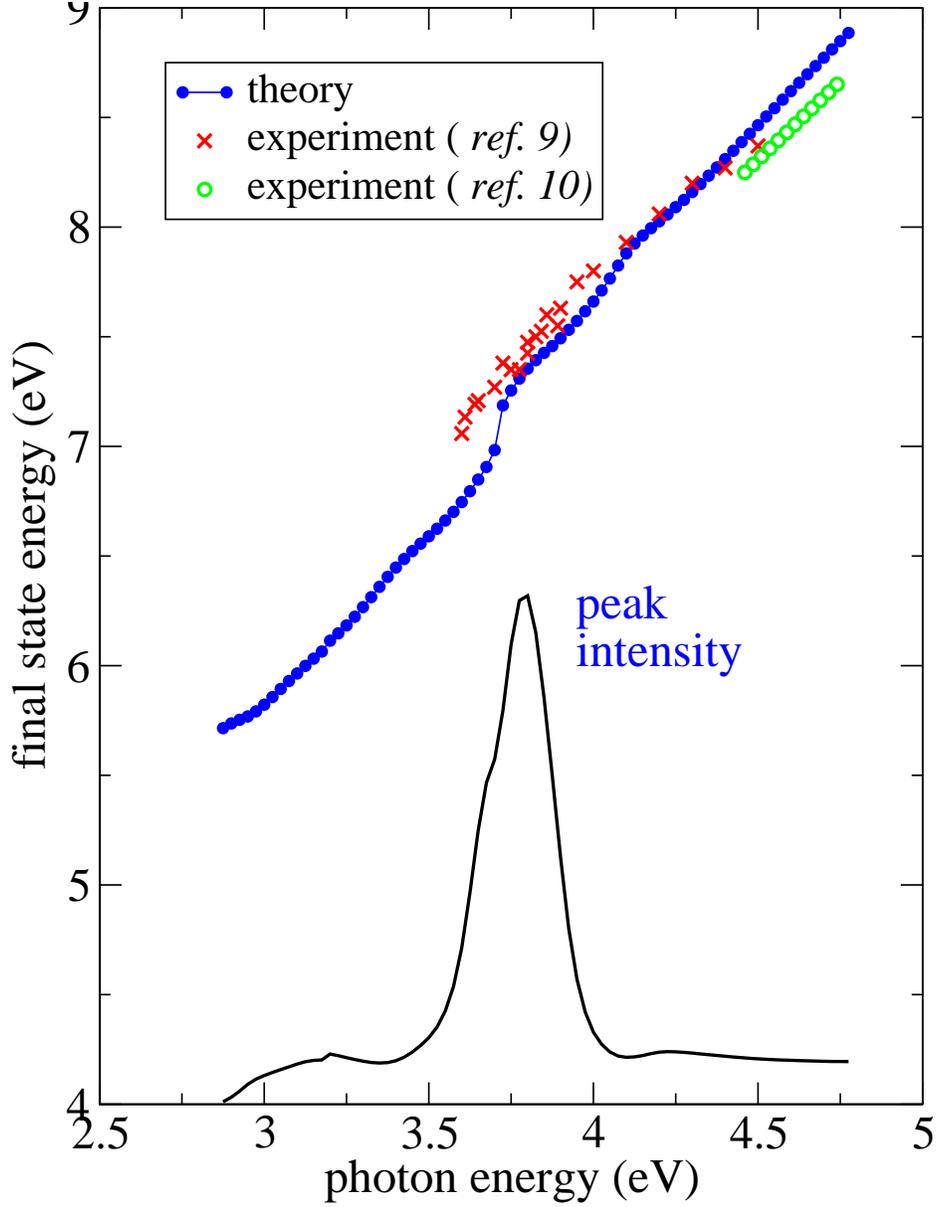}
\caption{(Color online) Peak magnitude and dispersion in the
two-photon spectra of Si(001). Calculated (full blue circles) peak
positions are shown as a function of photon energy. Experimental
values of Ref.~\cite{Shudo} (red crosses) and Ref.~\cite{Kentsch}
(open green circles) are shown as well. Solid black curve shows the
theoretical peak height in arbitrary units.} \label{sidispersion}
\end{figure}

The calculated peak magnitude and dispersion are shown in
Fig.~\ref{sidispersion} and compared to the corresponding measured
magnitudes \cite{Shudo, Kentsch}, finding good agreement as well.
The energy analysis shows an almost linear final state energy vs.
photon energy behavior. A kink is predicted at the photon energy for
which the maximum peak intensity is found. This is precisely the
energy at which the LEED state starts to play a role in the
two-photon photocurrent.

In addition to laser-based techniques, other standard photoemission
techniques, such as ARPES, can benefit from the analysis proposed in
this Letter. For illustration purposes, we show in the following a
model calculation on Al(111), in which photoemission spectra are
obtained from a similar one-step calculation. We represent the
Al(111) surface by a one-dimensional model in which the initial and
intermediate states are tight binding orbitals and the final states
are nearly free electron waves.  Only one branch of the final state
bands, which correspond to the LEED complex band structure that
allows electron escape, is considered. We also use this system to
discuss the expected magnitude of the non-linear peaks, as compared
to the one-photon peaks.

\begin{figure}[hb]
\psfig{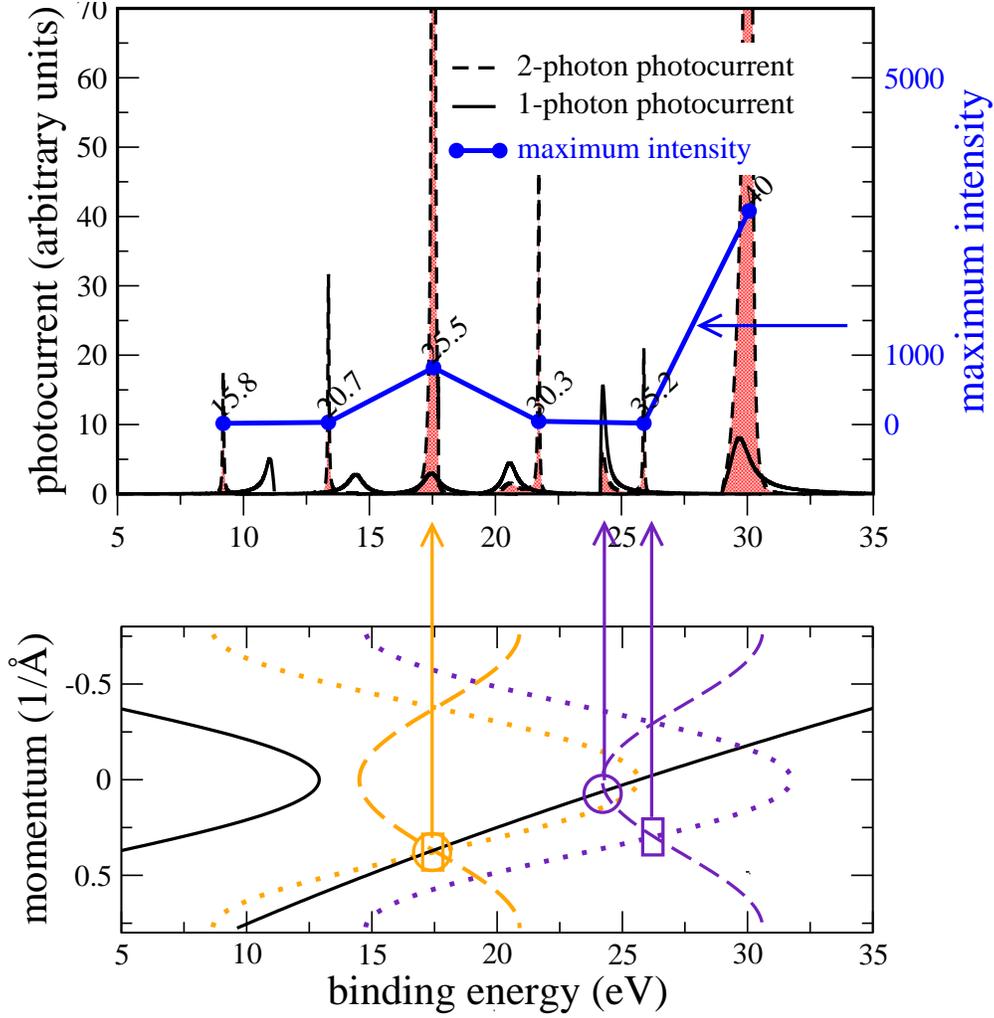} \caption{(Color online)  Upper panel:
Two-photon (broken lines) and one-photon (solid lines) photocurrents
for Al(111) (left ordinate), vs. binding energy. The maximum
intensity values for the two-photon current are also plotted as blue
circles (right ordinate), linked with a line to guide the eye.
Numbers over these circles denote twice the photon energy, in units
of eV. One-photon transitions are associated with doubled photon
frequency. Lower panel: Momentum vs. binding energy of model band
structure (solid lines). Two particular cases [$2\hbar\omega$ = 25.5
(orange) and 35.2 eV (purple)] are associated by arrows with the
corresponding peaks in the spectra above. The initial band has been
shifted by $2\hbar\omega$ (broken lines) and the intermediate state
band has been shifted by $1\hbar\omega$ (dotted lines). Rectangles
(circles) denote conservation of momentum between initial and
intermediate (initial and final) states.} \label{full}
\end{figure}

Figure \ref{full} shows both the two-photon photocurrent (with
$\eta=0.1$) and the one-photon current of doubled frequency. The
final state is the same for both types of photoemission. In each
two-photon spectrum, a very narrow peak appears and dominates the
spectrum by orders of magnitude. It comes from rather sharp
transitions from the initial to the intermediate state (denoted by
rectangles), given by the Lorentzian function in Eq.~(\ref{prob0}),
which has the same shape as an exactly direct transition at a
definite momentum. The intermediate to final state transition is
often obscured by the non-conservation of perpendicular momentum.
If, by chance, it is conserved too, a resonant magnification of the
peak occurs. In Fig.~\ref{full}, this happens for
$2\hbar\omega=25.5$ eV and $2\hbar\omega=40$ eV.

Furthermore, in the example of the $2 \hbar\omega = 35.2$ eV curve
in Fig.~\ref{full}, the rectangle shows the main peak resonance
occurring at a momentum different from the usual one-photon direct
transition into the final state parabola. As a consequence, the
current is significantly reduced as compared to the transitions that
fully conserve energy and momentum. A side peak arises at conserved
momentum between the initial and final states (denoted by circles),
as in one-photon spectra. The intermediate energy is not at its
resonance, however. Without exceptions, the initial and final
energies are sharp because of the $\delta$-function in
Eq.~(\ref{prob0}).

Summarizing, we propose a new methodology to extract
momentum-resolved band-structure information for unoccupied electron
states in bulk. The non-linear contributions of the two-photon
photoemission experimental spectra are used for this purpose. We
show that the composition of a bulk two-photon spectrum displays the
clear fingerprint of a Lorentzian line at exact momentum
conservation for the transition between initial and intermediate
state. It allows a direct association of both contributing bands and
the determination of their energy difference at that specific
momentum if the bands are unique. A standard Lorentzian
deconvolution is necessary in the case of multiple bands.
Experimental results for low photon energies on Si(001) support the
theoretical findings. With respect to explore and utilize these
effects in ARPES, the counter-play between higher light intensity
(higher photocurrent) and the concomitant broadening of the initial
to intermediate state transition by ballistic acceleration has to be
considered. Furthermore, a dispersion of the initial band sufficient
to fix the momentum of that transition is needed.

Within this methodology, the values of fundamental band gaps in
semiconductors appear as direct results of the two-photon spectra.
These results represent the optical direct band gaps at any momentum
where the transition via an intermediate state is observed. They are
in contrast with band gaps of direct and inverse photoemission where
the particle number is not conserved. Within the same reasoning, the
scissors operator, which rather artificially separates energetically
excited bands from the ground state, can be quantified. In Si(001),
for instance, the shift of excited bands to higher energies by the
real part of the self-energy could be adjusted to experiment as 0.3
eV. This shift seems to account for both a quasiparticle shift
upwards and an excitonic shift downwards.

Let us finally remark that non-linear optical processes in atoms and
molecules have been crucial in the discovery and understanding of
new physical phenomena in recent years. A similar burst for
condensed matter systems can be envisaged in the years to come. We
hope that the novel resonant processes proposed in this Letter can
serve as an additional motivation to theorists and experimentalists
alike to further explore the exciting capabilities of non-linear
spectroscopies in solids.

We are grateful to Hrvoje Petek for a critical reading of the
manuscript. Financial support by DFG under contract FOR353 and the
Spanish MEC (Grant No. FIS2007-66711-C02-02) is acknowledged.

\newpage
%
%

\newpage
%
%

\newpage
%
%

\newpage

%
%


\begin{thebibliography}{55}
\bibitem{footnote} In one-photon photoemission, the uncertainty in
perpendicular momentum $\Delta k_{\perp}$ can be estimated from the
optical potential $V_i$ as $\Delta k_{\perp}=(2V_{i})^{1/2}$,
whereas in two-photon photoemission the only limits to the full
conservation of momentum are those linked to the optical skin depth.
\end{thebibliography}
\end{document}